\documentclass[a4paper]{jpconf}
\usepackage{graphicx}
\begin{document}
\title{High Energy Neutrinos: Sources and Fluxes}
 
\author{Todor Stanev}

\address{Bartol Research Institute, Department of Physics and Astronomy, University of Delaware, Newark DE 19716, U.S.A.}

\ead{stanev@bartol.udel.edu}

\begin{abstract}
  We discuss briefly the potential sources of high energy astrophysical
 neutrinos and show estimates of the neutrino fluxes that they can
 produce. A special attention is paid to the connection between
 the highest energy cosmic rays and astrophysical neutrinos.
\end{abstract}

\section{Introduction}
 High energy is defined here as that of neutrinos above
 atmospheric neutrino background, which is generated in 
 cascades induced by cosmic rays in the atmosphere. The production
 mechanism is the decay chain of charged mesons and muons, i.e.
 $\pi^{\pm} \rightarrow \nu_\mu (\bar{\nu}_\mu) + \mu^{\pm}$,
 $\mu^{\pm} \rightarrow \bar{\nu}_\mu (\nu_\mu) + \nu_e (\bar{\nu}_e) +
 e^{\pm}$ and the respective chain of K decays.

  Neutrinos from meson decay have energy
 spectrum steeper than that  of the cosmic rays by one power of E.
 Neutrinos from muon decay are 
 steeper by another power of E. Because of the steep atmospheric
 neutrino energy spectrum (E$_\nu^{3.7 - 4.7}$) we expect to start
 seeing astrophysical neutrinos at energies above 1 TeV. That
 threshold is somewhat uncertain because of the contribution of direct
 neutrinos from decays of charm and heavier flavors.
 Such neutrinos are expected
 to cross the meson decay neutrino spectrum at energies between
 10 and 100 TeV.

 There is an important question that we have to answer
 first: what is the intellectual value of the detection of astrophysical
 neutrinos compared to the detection of astrophysical TeV $\gamma$-rays,
 which is experimentally much easier. The reply to this question is 
 even more important now after the extraordinary success of HESS and 
 the incoming results of the Magic Cherenkov telescope~\cite{here1}. 

 The answer is that the Universe is not transparent to high energy
 $\gamma$-rays. Even in the absence of matter and local photon
 fields PeV $\gamma$-rays are absorbed in pair production 
 interactions on the microwave background (MBR). TeV $\gamma$-rays are 
 absorbed in the infrared/optical 
 background (IRB). So one can have hidden sources of astrophysical
 neutrinos, either surrounded by large column densities of matter,
 or intense photon fields, or just being far away from us.         

 Classical high energy astrophysics is interpreting all
 processes as electromagnetic. Electromagnetic models have been very
 successful in describing the production of TeV $\gamma$-rays in 
 different astrophysical objects. Only the detection of astrophysical
 neutrinos can point at the existence of hadronic processes at
 astrophysical objects.

 There are two types of hadronic processes that lead 
 to the production of mesons and the start of the meson
 and muon decay chains: inelastic hadronic interactions
 on matter (which we shall refer to as $pp$ interactions)
 and photoproduction interactions on photon fields ($p \gamma$
 interactions). Apart from the  interaction energy threshold,
 where the main process is the $\Delta^+$ production with
 a cross section of 500 $\mu$b, the photoproduction  cross
 section is smaller than the $pp$ inelastic cross section by two
 orders of magnitude. Since the microwave background density is
 higher than 400 cm$^{-3}$, photoproduction interaction is
 four times more likely when the matter density is 1 cm$^{-3}$.
 On the other hand the energy threshold for photoproduction
 is high because of the very low energy of the background
 photons. One way to  express the minimum proton energy is
 E$_p^{min}\; = \; (m^2_{\Delta} - m^2_p)/2 \varepsilon$, where 
 $\varepsilon$ is the energy of the background photon.

 Traditionally we use to think of inelastic proton interactions in
 galactic astrophysical systems, such as supernova remnants,
 and of  photoproduction interactions in powerful extragalactic
 objects~\cite{GHS95}. These ideas are in the process of changing
 now, when the similarities of many galactic and extragalactic
 astrophysical systems start to emerge. We believe that in astrophysical 
 environment all mesons and muons decay and the neutrino spectra are
 thus parallel to the meson (and cosmic ray) spectra. The only possible
 spectrum modification comes from energy loss of the mesons and,
 most likely, muons.

\section{Galactic neutrino sources}

 There are several types of suspected galactic sources of astrophysical
 neutrinos. These include supernova remnants, powerful binary systems,
 microquasars, and the Galactic center.

  Generally we expect to see neutrinos from any object that generates 
 $\gamma$-rays of hadronic origin ($\pi^0 \rightarrow \gamma\gamma$).
 The generic rule is that if an object produces $\gamma$-rays on a 
 A$E_\gamma^{-\alpha}$ spectrum from $\pi^0$ decay it will generate
 neutrinos from $\pi^\pm$ decay on a spectrum
 A$(1-r_\pi)^{\alpha-1}E_\nu^{-\alpha}$, where $r_\pi = (m_\mu/m_\pi)^2$.
 When one also accounts
 for the neutrinos from muon decay for flat injection spectra the
 neutrino flux is almost equal to the $\gamma$-ray flux. 
 
  Microquasars are mini versions of Active Galactic Nuclei that are
 respectively less powerful, but much closer to us. We shall
 discuss the neutrino production at AGN in the following section

  With its great power in any other wavelengths SGR A$^*$ could be 
 also a neutrino source, especially after the detection of TeV
 $\gamma$-rays from that source~\cite{HESS_SGRA}. The detected 
 $\gamma$-ray emission extends up to 10 TeV on a flat E$_\gamma^{-2.2}$
 energy spectrum. See also the talk of D.~Grasso at this meeting.

  Before we briefly discuss the neutrino production at supernova
 remnants and binary systems, we want to introduce two guaranteed
 sources of neutrinos, neither of them of high astrophysical interest.

 The first one is the Sun. The atmosphere of the Sun is
 more rarefied and thus the generated neutrinos reach much
 higher energy than atmospheric ones~\cite{ssg91}. Thus at energies
 exceeding 1 TeV the Sun becomes a neutrino source that could be
 used for calibration of the neutrino telescopes.

  Another guaranteed source is the central region of the Galaxy, 
 where the EGRET group has measured diffuse GeV $\gamma$-ray 
 emission with a spectral index of 2.3-2.4~\cite{EGRET-CG}.
 The emission could indeed be diffuse, or it could be due to
 unresolved sources. Recently the HESS collaboration surveyed
 the central Galaxy and discovered 14 individual $\gamma$-ray 
 sources extending upwards from 200 GeV~\cite{HESS-CG}. If all
 these sources were hadronic, the generated neutrino emission 
 would easily be detectable by km$^3$ neutrino telescopes.

\subsection{Neutrino production at supernova remnants} 
  
 There are two periods when a supernova can be a source
 of hadronic $\gamma$-rays and neutrinos. The production of
 signals at young supernova remnants is discussed
 in detail for $\gamma$-rays by Berezinsky\&Prilutski~\cite{BerPri}.
 The emission
 starts whenever protons can be accelerated and mesons start 
 decaying rather then interacting and lasts while the target 
 density in the expanding supernova shell is sufficiently high.
 Depending on the supernova parameter, the emission may last
 between one and three years. A modern discussion of the
 processes in young supernova like objects can be found 
 in Ref.~\cite{WaxMesz} who consider the emission from hypernova
 explosions in M82 and NGC253. These {\em collapsar} neutrinos
 do not present a significant diffuse flux, but could be 
 detectable by the coincidence with the hypernova event.

 The other period of high energy signal activity is when the
 supernova envelope drags out enough interstellar matter to
 slow it down - during the Sedov phase, about 1000 years after
 the supernova.
 During that phase the lost kinetic energy of the remnant 
 is converted into cosmic rays, and the matter density at the
 blast shock becomes higher~\cite{DAV94}. It is even more 
 likely that signals are produced when the expanding supernova
 remnant hits a dense molecular cloud~\cite{ADV94}. The same
 conclusion was reached in an analysis~\cite{GPS98} of the 
 emission detected by the EGRET instrument from the vicinity
 of supernova remnants.
 Hadronic $\gamma$-ray production dominates over
 electromagnetic one when the matter density exceeds 100 cm$^{-3}$.
 This is the case of the remnant RXJ 1713.7-3946 that was 
 observed by the HESS telescope~\cite{HESS_Nat}.
 HESS has a fine angular resolution and studied the strength of the
 $\gamma$-ray signal from different regions of the remnant. It is well 
 known that  RXJ 1713.7-3946 is hitting a molecular cloud of density
 higher than 100. It is thus possible that at least a fraction of 
 the detected $\gamma$-ray flux of 1.7$\times$10$^{-7}$ E$_{TeV}^{-2.2}$
 cm$^{-2}$s$^{-1}$TeV$^{-1}$ has $\pi^0$ origin and we can expect 
 from the object a neutrino flux of the same order.  

 It is worth noting that the supernova remnant SGR A East is
 in the error box of the HESS detection of SGR A$^*$ and it
 can not be proven that the detected TeV $\gamma$-ray emission
 does not come from this supernova remnant.

\subsection{Neutrino production at binary systems}

 Powerful binary systems offer a large number of possibilities 
 for particle acceleration and interactions. 
 The emission from the compact object runs into the stellar wind
 of the companion star and shocks can be easily formed.
 The stellar wind provides enough nucleons for 
 particle acceleration and the matter density of the star, 
 the wind itself, and the accretion wake provide targets 
 for interaction of the accelerated particles~\cite{BP79}. 

 The observations of the central Galaxy by HESS~\cite{HESS-CG}
 identified several binary systems as sources of TeV $\gamma$-rays.
 In the near future we will see if these binary systems are 
 also sources of astrophysical neutrinos.

\section{Extragalactic sources}

 The extragalactic sources that are usually considered are 
 active galactic nuclei (AGN) and gamma ray bursts (GRB).
 Because of their high luminosity at other wavelengths,
 AGNs have been long suspected~\cite{bz77} as possible sources
 of astrophysical neutrinos. There are two possibilities
 for particle acceleration and interactions in AGN: the
 central region close to the black hole, and the AGN jet.

\subsection{Generic AGN}

 The neutrino production in the central region of AGN has been
 first discussed in Refs.~\cite{SDSS,SP94}. Both papers calculate
 the  accretion rate that is needed to support the black hole
 luminosity which gives them the nucleon density in the
 vicinity of the black hole. The photon density is estimated
 directly from the luminosity of the object. It turns out
 that the photon density exceeds the nucleon one by
 many orders of magnitude. The main
 energy loss of the accelerated protons will than be in 
 photoproduction interactions that generate a  
 neutrino flux of specific shape, that follows the proton
 spectrum above the
 interaction threshold and is flat below it.

 The problem with these models is that they are very 
 difficult to normalize since no other type of radiation 
 can directly be observed - it is absorbed and downgraded 
 in energy by the intense radiation field. One can only guess
 what fraction of the MeV/GeV isotropic background could be 
 produced in such generic AGN. 

\subsection{AGN jets}

 The signals produced in the AGN jets are boosted by the Doppler
 factor of the jet which could be as high as 10. The detected
 TeV $\gamma$-ray emission from BL Lac objects is explained well
 with inverse Compton scattering of accelerated electrons.
 These models reproduce the double peaked distribution of the
 jet photon emission and can accommodate the fast variability
 of the sources.

 Hadronic models of the jet emission were first developed by
 Mannheim\cite{Mannheim}. These models can also reproduce the
 double peaked spectra, although the variability is more difficult
 to account for.
 It is quite possible that the $\gamma$-ray and neutrino emission in
 jets are not proportional to each other~\cite{AMuecke03}. 
 Under different astrophysical assumptions either strong $\gamma$-ray
 or strong neutrino signals are produced.

 The basic principles of the proton acceleration and interactions
 are discussed in Ref.~\cite{RM}. These are equally valid for
 AGN jets and gamma ray bursts, where the physics is similar to
 this in jets, but the conditions are more extreme.

\subsection{Gamma ray bursts}

 The neutrino emission from GRB was discussed after these objects
 were suspected to be the sources of the highest energy cosmic
 rays~\cite{GRB_CR}. The extremely high GRB luminosity and
 the average GRB Lorentz factor of 300 make the production of 
 high energy neutrinos at these objects quite possible. 
 The observed GRB photon spectrum has a break in the vicinity
 of 1 MeV, which in turn defines a neutrino spectrum with a 
 break at about 10$^{14}$ eV~\cite{WB97}.

 Once again, the photon emission and neutrino emission may not
 be proportional to each other~\cite{HHAHH}. Fast GRB jets with
 a Lorentz factor of up to 1,000 are very good photon emitters,
 while slow jets with a Lorentz factor around and below 100 may
 be very good neutrino emitters. The same conclusion is reached 
 in a detailed calculation~\cite{DA03}, where the neutrino 
 production is discussed for GRB models with internal and external
 soft photon origin.

 The best ones are stalled GRB~\cite{MW01},
 where the jet is not able to penetrate through the collapsing
 supermassive star. In such case, however, the main 
 advantage of the GRB neutrino observation - time coincidences
 with the explosion - will be lost. 

\section{High energy neutrino fluxes and ultrahigh energy cosmic rays}

 Waxman\&Bahcall
 did a simple calculation of the production of neutrinos
 at optically thin sources of UHECR~\cite{WBub}, assuming
 that  the total energy of the accelerated protons is converted to 
 neutrinos. The calculation uses the emissivity that can maintain 
 the flux of UHECR and gives an upper bound of the extragalactic
 neutrino flux, expressed as $E_\nu^2 dN_\nu/dE_\nu = 5\times 10^{-8}$
 GeV cm$^{-2}$s$^{-1}$ in the case of $(1+z)^3$ cosmological evolution
 of the cosmic ray sources. In terms of energy flux this is a straight 
 line extending up to energies of 10$^{20}$ eV.

 This bound was discussed and somewhat relaxed in Ref.~\cite{MPR}
 where are a more sophisticated calculation was performed. The 
 two bounds agree with each other only at E$_\nu$ of 10$^{18}$ eV.

 The upper bounds cannot  be accepted as a 
 upper limit on the astrophysical neutrino fluxes. It does not
 restrict, e.g. the fluxes from sources that are optically thick
 for nucleons and is exceeded by 
 fluxes of cosmogenic neutrinos.
 
\subsection{Cosmogenic neutrinos}

 These are neutrinos generated by photoproduction interactions
 of the extragalactic cosmic rays in the photon background.
 The most important universal target is the microwave background. 
 The existence of such neutrinos was first proposed in Ref.~\cite{BerZat69}
 and developed in~\cite{Stecker73}. The relation of this flux to the 
 cosmological evolution of the cosmic ray sources was discussed
 in~\cite{HS85}. The fluxes of cosmogenic neutrinos were 
 more recently calculated in many other papers.

 There are several important astrophysical inputs in such a
 calculation, the most important among which are
 {\em 1)} the emissivity of the cosmic ray sources, estimated from the
 the flux of the UHECR, {\em 2)} the distribution of the UHECR sources,
 which is usually assumed to be isotropic and homogeneous,
 {\em 3)} the cosmological evolution of the cosmic rays sources, 
 usually set parallel to this of star forming regions, and
 {\em 4)} the cosmic ray injection spectrum.
 
 In the current Universe, at redshift $z=0$, the proton threshold
 energy for photoproduction interactions on MBR is about
 3$\times$10$^{19}$ eV. The emissivity normalization is then
 set to an energy where one believes all cosmic rays are
 extragalactic. For cosmological evolution
 of the cosmic ray sources as $(1+z)^3$ Waxman (1995) calculates
 the cosmic ray emissivity above 10$^{19}$ eV to be
 4.5$\times$10$^{44}$ erg/Mpc$^3$/yr for a wide range of injection
 spectra. 

 The resulting diffuse flux~\cite{ESS01} of $\mu_\mu + \bar{\nu}_\mu$
 peaks  at about 5$\times$10$^{17}$ eV and reaches $dN/d(lnE)$ of
 3$\times$10$^{-17}$ cm$^{-2}$s$^{-1}$ster$^{-1}$ at the peak for
 E$^{-2}$ injection spectrum of UHECR and $(1+z)^3$ cosmological
 evolution of the cosmic ray sources.
 At higher energy the neutrino flux roughly follows the UHECR 
 spectrum, while below 10$^{16}$ eV the flux us flat. 

 Since only protons of energy above 3$\times$10$^{19}$ eV interact
 on MBR the flux is lower for steeper injection spectra. If they were
 no cosmological evolution of the cosmic ray sources the peak moves
 to higher energy and the peak flux is lower by about a factor of 5.
 The relation between the cosmic ray injection power law index $\gamma$
 and the cosmological evolution parameter $m$ from the $(1+z)^m$ 
 dependence can be expressed as $(1+z)^{(m+\gamma-\frac{3}{2})}$ when
 the redshift integration is performed over $ln{1+z}$~\cite{SS05}. 
 For $(m + \gamma)$ greater than $\frac{3}{2}$ the production 
 increases with redshift. Otherwise it decreases.

 Most of the cosmogenic neutrino calculations are performed in the
 assumption that UHECR are protons. It was recently demonstrated
 \cite{Aveetal05} that there would be detectable flux of 
 cosmogenic neutrinos even if UHECR were heavy nuclei. The 
 neutrinos then peak at about 10$^{14}$ eV ($\bar{\nu}_e$) from
 neutron decay, although there is also a secondary peak at
 higher energy, similar to the one shown in Fig.~\ref{yieldfl}.

\subsection{Cosmogenic neutrinos from interactions on the infrared
 background}

 The infrared and optical background is another universal
 photon field that has been studied directly and through the
 absorption of the TeV $\gamma$-rays from distant systems.
 The IRB number density is smaller than that of MBR by factors
 from 250 to 400 in different models, but the important factor
 is that the proton threshold energy on interactions on IRB is
 about 10$^{17}$ eV. The increased number of lower energy
 protons to a large extend compensates for the lower target
 density of IRB. The important region of IRB is the far and
 medium infrared regions where the number density is much higher 
 than in the near IFR and optical range.

\begin{figure}[htb]
\includegraphics[width=0.495\textwidth]{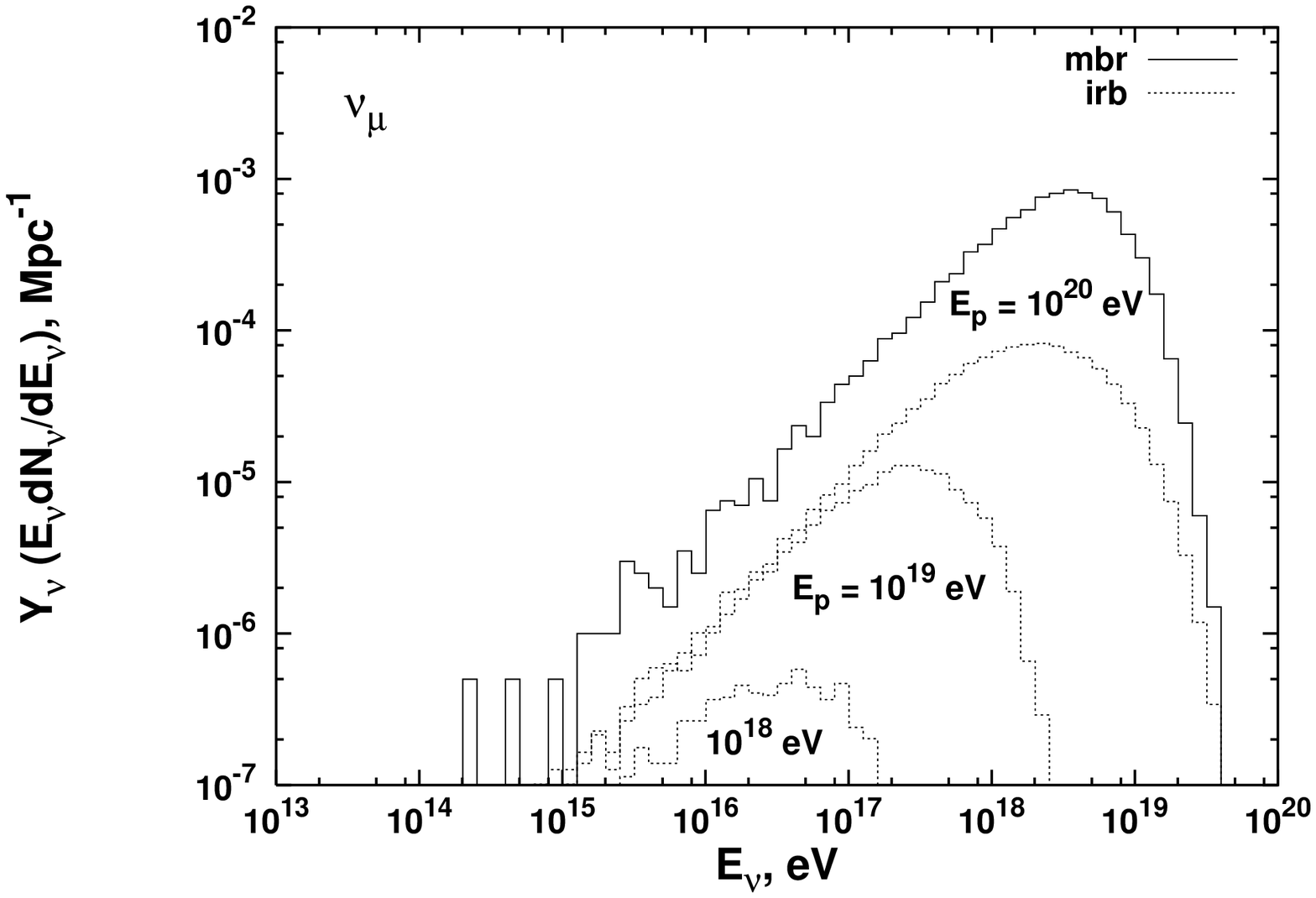}
\includegraphics[width=0.495\textwidth]{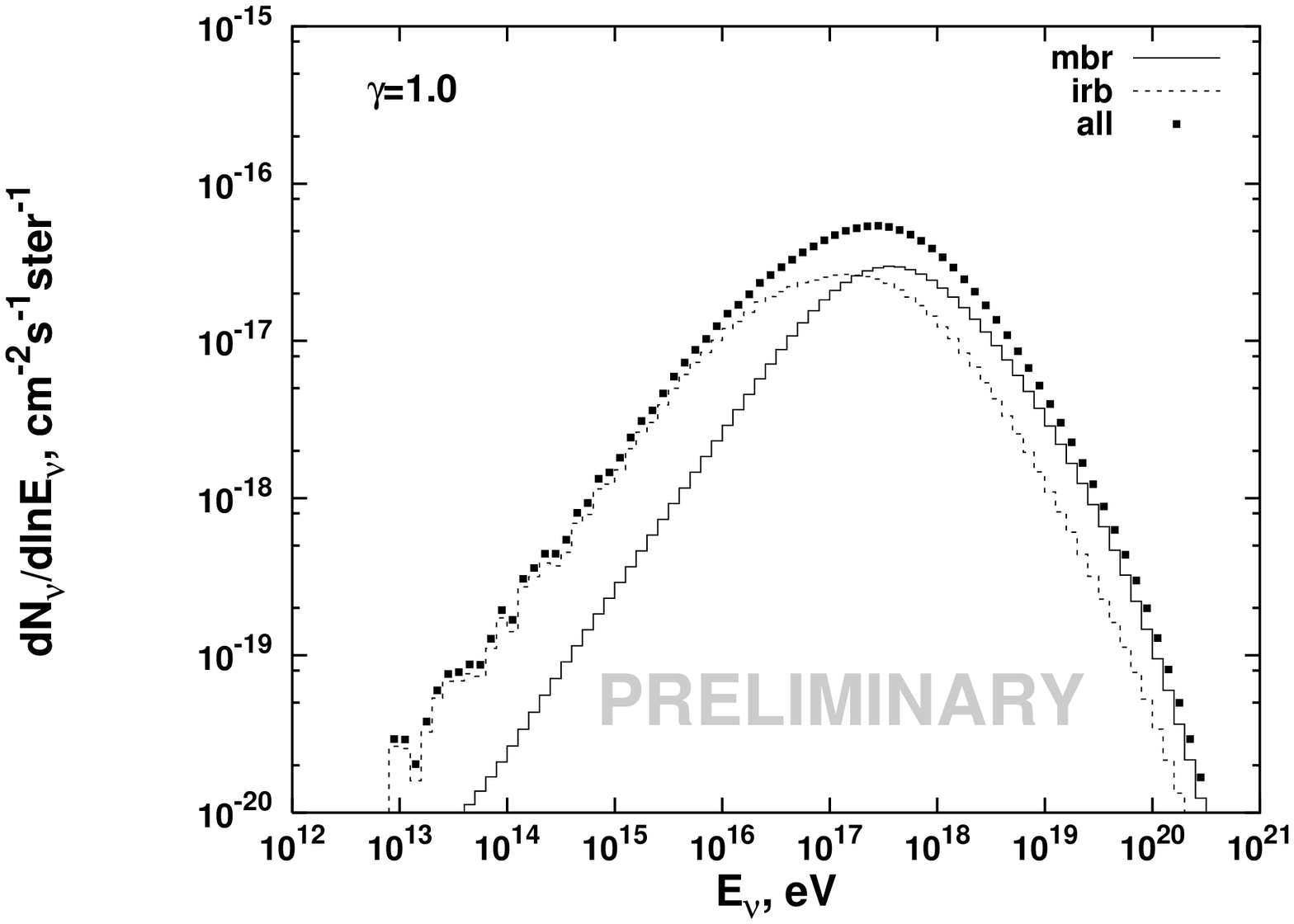}
\caption{ Left-hand panel: muon neutrino yields from proton propagation
 on 1 Mpc in the MBR (solid) and IRB (dotted). Proton energies are
 indicated by the histograms. Right-hand panel: cosmogenic neutrinos
 from interactions on MBR (solid) and on IRB (dots). The total 
 flux is shown with points.}
\label{yieldfl}
\end{figure}

 Fig.~\ref{yieldfl} shows in its left panel the yields from 
 interactions on the MBR and on IRB for propagation on distance
 of 1 Mpc. The photoproduction interactions are modeled with
 the SOPHIA code~\cite{SOPHIA}.
 Although the IRB yield for proton energy of 10$^{20}$ eV
 is much lower than that on MBR, there are also yields from
 10$^{19}$ and 10$^{18}$ eV protons. 

 The right-hand panel of Fig.~\ref{yieldfl} shows the cosmogenic
 neutrinos generated on MBR and on the IRB from the model of
 Ref.~\cite{SMS}.
 The calculation uses the cosmological evolution of IRB as calculated
 in this model up to redshift of 5. Waxman's cosmic ray emissivity is
 used together with injection spectrum $\gamma$=1 and cosmological 
 evolution $(1+z)^3$ of the cosmic ray sources.

 The total flux of cosmogenic neutrinos consists of roughly equal 
 parts of MBR and IRB interaction neutrinos. The IRB flux covers
 wider energy range and is slightly shifted to lower energy.
 The decrease of the number of high energy neutrinos in IRB 
 interactions can be explained with the fact that UHECR protons
 interact mainly on the MBR.

 The picture changes for steeper cosmic ray injection spectra.
 The number of lower energy protons grows faster, and so does
 the contribution of the IRB interactions. For $\gamma$=1.5,
 for example, the IRB peak is higher than the MBR one by more
 than a factor of 3 and the total neutrino energy distribution
 is shifted to lower energy
 
 The IRB contribution thus decreases the dependence of the cosmogenic
 neutrino flux on the cosmic ray injection spectrum. While the
 MBR neutrino flux decreases for steeper injection spectrum, the
 IRB flux rapidly increases. Accounting for the total photon
 field makes the cosmogenic neutrino flux roughly independent of
 the proton injection spectrum.

\section{High energy neutrino fluxes} 

\begin{figure}[htb]
\includegraphics[width=0.495\textwidth]{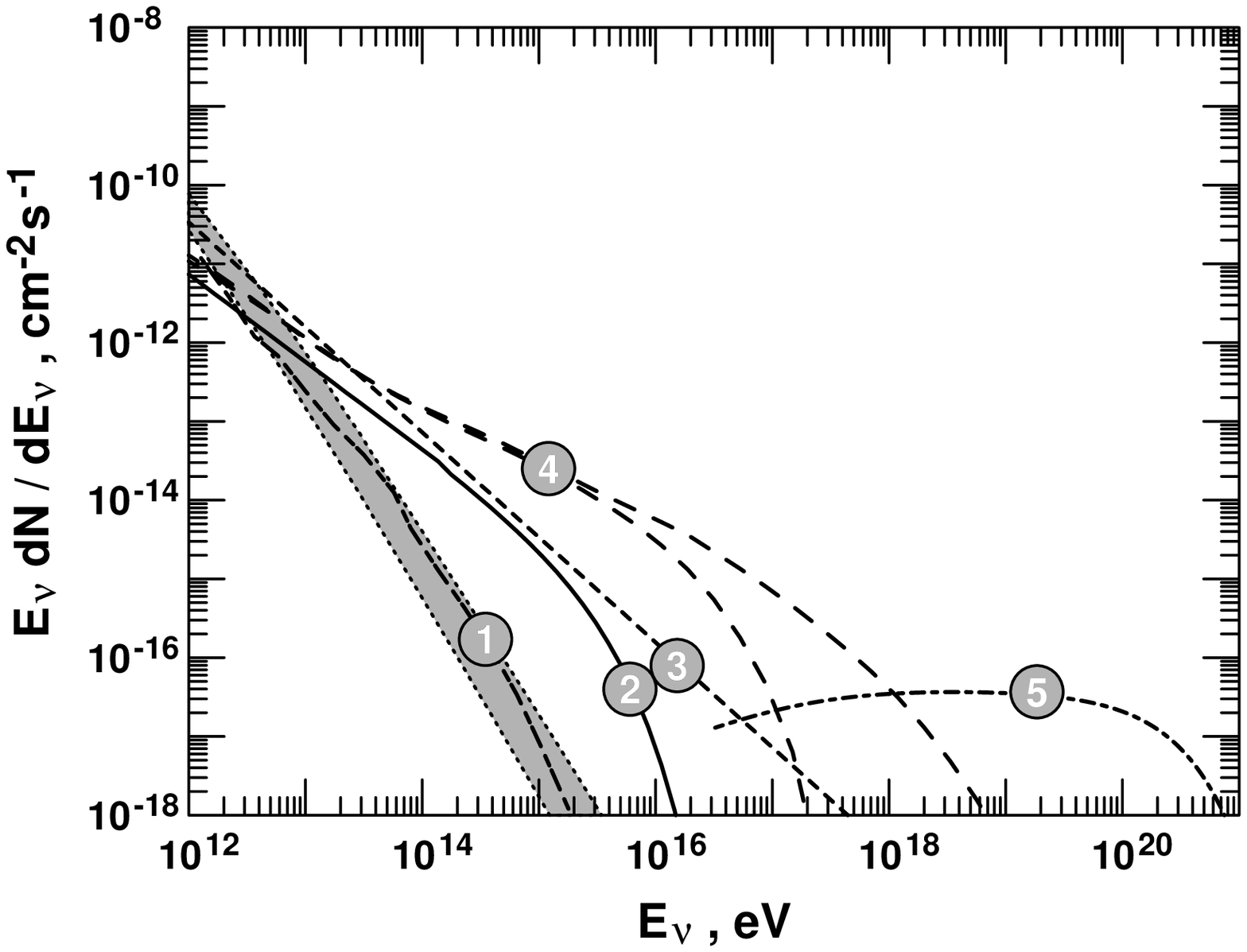}
\includegraphics[width=0.495\textwidth]{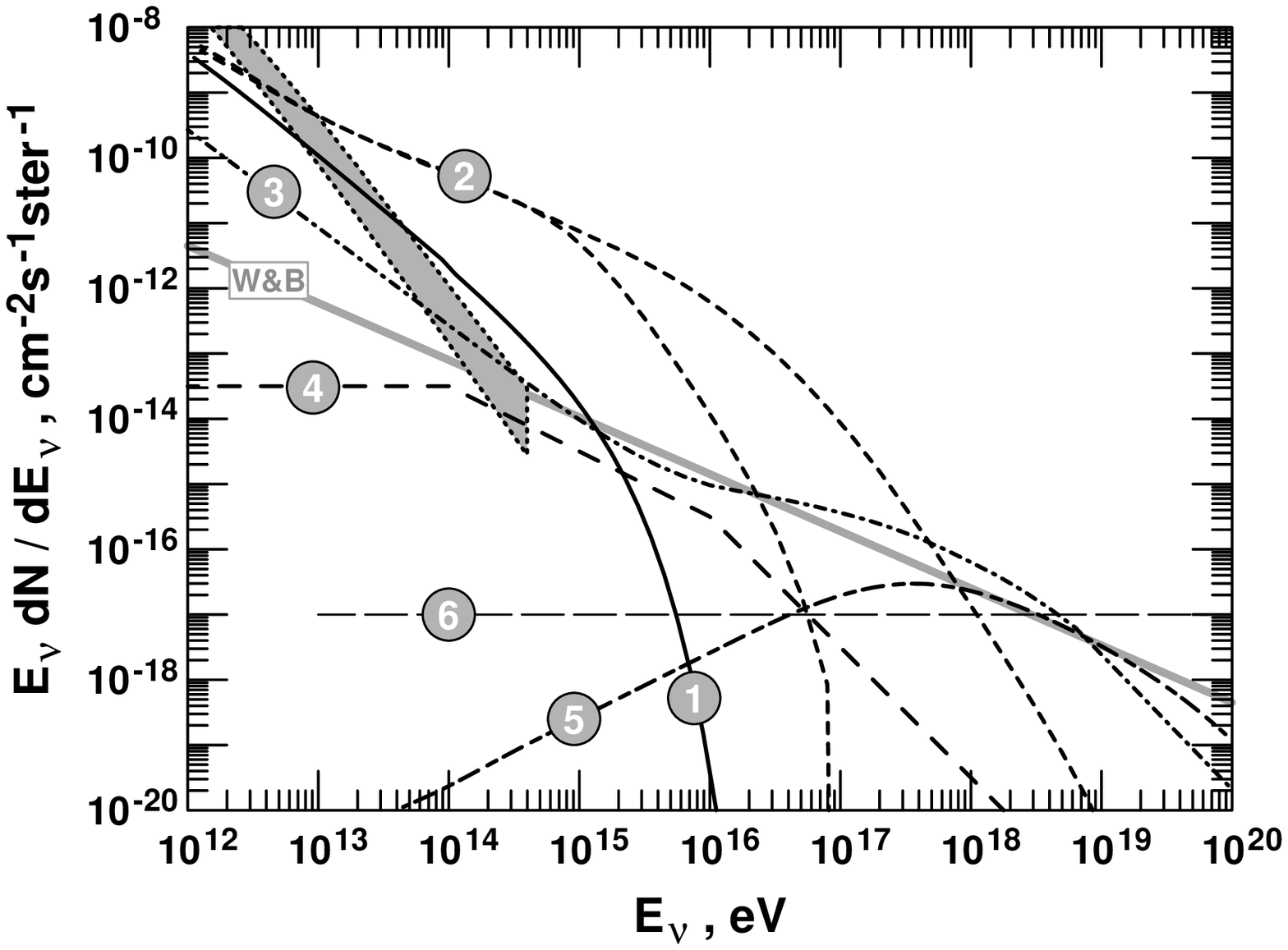}
\caption{ Left-hand panel: fluxes from potential neutrino sources.
 Right-hand panel: diffuse neutrino fluxes from unresolved neutrino sources.
 }
\label{fluxes}
\end{figure}

 Left-hand panel of figure~\ref{fluxes} shows $\nu_\mu + \bar{\nu}_\mu$
 fluxes predicted for five potential neutrino sources. The
 atmospheric neutrino fluxes within 1$^\circ$ from the source are
 indicated with a shaded region. Upper end is for horizontal
 neutrinos and lower one - for vertical neutrinos.

 The curve labeled {\em 1)} shows the neutrinos that we expect from 
 the direction of the Sun~\cite{ssg91}.
 Curve {\em 2)} shows the neutrino fluxes expected from the
 supernova remnant IC443 if the $\gamma$--rays detected 
 by EGRET are all of hadronic origin~\cite{GPS98}.
 Curve {\em 3)} shows the expected neutrino luxes if the TeV
 $\gamma$-ray outburst  of Mrk 501 is of hadronic origin.
 Curve {\em 4)} shows the minimum and maximum fluxes expected
 from the core region of 3C273~\cite{SP94}.
 Curve {\em 5)} shows the neutrino flux predicted for the jet of
 3C279~\cite{Mannheim}.

 The right-hand panel of figure~\ref{fluxes} shows several different
 diffuse astrophysical neutrino fluxes. The shaded area indicates
 the vertical and horizontal fluxes of atmospheric neutrinos.
 Waxman\&Bahcall bound~\cite{WBub} is indicated with W\&B.

 The curve labeled {\em 1)} shows the neutrinos expected from the
 central Galaxy in the assumption that all diffuse $\gamma$-rays
 detected by EGRET~\cite{EGRET-CG} are created  by cosmic
 ray interactions with matter. Curves {\em 2)} are diffuse fluxes
 from cores of AGN~\cite{SP94}.
 Flux {\em 3)} is the
 isotropic AGN neutrino flux from AGN jets~\cite{Mannheim}, 
 where $pp$ interactions are added to the high energy photoproduction
 interactions. Flux {\em 4)} is the prediction of diffuse neutrinos from
 GRB~\cite{WB97} in the assumption that GRBs are sources of the
 ultrahigh energy cosmic rays.
 Flux {\em 5)} is a nominal cosmogenic neutrino flux as calculated in 
 Ref.~\cite{ESS01} using the luminosity and cosmological evolution
 model from the W\&B limit. 
 Finally, for comparisons with diffuse astrophysical
 neutrinos we show the neutrino flux {\em 6)} that is
 needed by the  Z-burst model to become the production
 mechanism for UHECR~\cite{SSigl04}.
  
\ack  This research is supported in part by the US Department
 of Energy contract DE-FG02 91ER 40626 and by NASA Grant NAG5-10919.

\end{document}